\documentclass[preprint,eqsecnum,aps,nofootinbib]{revtex4}
\usepackage{amsfonts,amsmath,amssymb,amsthm}
\usepackage{latexsym}
\usepackage{bbm,bm}
\usepackage{graphicx}

\newcommand{\beq}{\begin{equation}}
\newcommand{\eeq}{\end{equation}}
\newcommand{\beqs}{\begin{eqnarray}}
\newcommand{\eeqs}{\end{eqnarray}}

\begin{document}

\title{Sum Rule of Quantum Uncertainties: Coupled Harmonic Oscillator System with Time-Dependent Parameters}

\author{DaeKil Park$^{1,2}$ and Eylee Jung$^1$,}

\affiliation{$^1$Department of Electronic Engineering, Kyungnam University, Changwon
                 631-701, Korea    \\
             $^2$Department of Physics, Kyungnam University, Changwon
                  631-701, Korea    
                      }

\begin{abstract}
Uncertainties $(\Delta x)^2$ and $(\Delta p)^2$ are analytically derived in an $N$-coupled harmonic oscillator system when spring and coupling constants are arbitrarily time-dependent and each oscillator is in an arbitrary excited state. When $N = 2$, 
those uncertainties are shown as just arithmetic average of uncertainties of two single harmonic oscillators. We call this property as ``sum rule of quantum uncertainty''.  However, this arithmetic average property is not generally maintained when $N \geq 3$, but it is recovered in $N$-coupled oscillator systems if and only if $(N-1)$ quantum numbers are equal. 
The generalization of our results to a more general quantum system is briefly discussed. 
\end{abstract}

\maketitle

\section{Introduction}

Uncertainty\cite{uncertainty,kennard1927,robertson1929,Busch2007} and entanglement\cite{schrodinger-35,text,horodecki09} are two major cornerstones of quantum mechanics. These characteristics cause quantum mechanics to differ from classical mechanics. Quantum uncertainty provides a limit on the 
precision of measurement for incompatible observables. The most typical expression of uncertainty relation is $\Delta x \Delta p \geq \hbar / 2$, where $\Delta$
is the standard deviation. Recently, researchers have analyzed different expressions of uncertainty relations,  such as entropic uncertainty relations\cite{Wehner2010,Coles2017}
from the context of quantum information and generalized uncertainty principle\cite{GUP} from the context of Planck scale physics. Even though entanglement has been studied 
since the discovery of quantum mechanics\cite{schrodinger-35}, it has been extensively explored for the last few decades with the development of quantum technology. 
Entanglement is used as a physical resource in various quantum information processing, such as  quantum teleportation\cite{teleportation,Luo2019},
superdense coding\cite{superdense}, quantum cloning\cite{clon}, quantum cryptography\cite{cryptography,cryptography2}, quantum
metrology\cite{metro17}, and quantum computer\cite{qcreview,computer}. Furthermore, with many researchers trying to realize such quantum information processing in the laboratory for the last few decades, quantum cryptography and quantum computer seems to approaching the commercial level\cite{white,ibm}.

Although these two phenomena seem to be distinct properties of quantum mechanics, there is some connection, albeit unclear, between them because of the fact that both are strongly dependent on the  interaction between subsystems. For example, the uncertainty of a given system was computed in Ref.\cite{han97,han99} to discuss on the effect of the Feynman's rest of universe\cite{feyman72}. The ignoring of the effect of the rest of the universe was shown to increase
uncertainty and entropy in the target system (system in which we are interested). In other words, if the target system is one of subsystems of a whole system and it interacts with other subsystems, its 
uncertainty and entanglement monotonically increase with increasing interaction strength. More specifically, let us consider  two coupled harmonic
oscillator system, with the following Hamiltonian:
\begin{equation}
\label{hamil-2}
H_2 = \frac{1}{2} \left( p_1^2 + p_2^2 \right) + \frac{1}{2} \left[k_0 (x_1^2 + x_2^2) + J (x_1 - x_2)^2 \right].
\end{equation}
If we assume that the two oscillators,  say $A$ and $B$, were in each ground state, the uncertainty and entanglement of 
formation (EoF)\cite{benn96} are both given by\cite{park18}
\begin{equation}
\label{two-vacuum}
\left( \Delta x \Delta p \right)_{A,B}^2 = \frac{1}{4} \left( \frac{1 + \xi}{1 - \xi} \right)^2       \hspace{1.0cm}
{\cal E}_F = - \ln (1 - \xi) - \frac{\xi}{1 - \xi} \ln \xi
\end{equation}
where $\hbar = 1$ and $\xi = \left\{(\sqrt{k_0 + 2 J} - \sqrt{k_0}) / (\sqrt{k_0 + 2 J} + \sqrt{k_0}) \right\}^2$. This shows that both 
$\left( \Delta x \Delta p \right)_{A,B}^2$ and ${\cal E}_F$ increase with increasing the coupling constant $J$. Thus, in this case, uncertainty and entanglement are implicitly related to each other via $\xi$. Duan et al.\cite{criterion} used quantum uncertainty to provide a sufficient criterion for entanglement in continuous variable systems. 
 Mandilara and Cerf\cite{cerf12} showed that the uncertainty relation for all eigenstates in the single harmonic oscillator system is saturated with respect
to Gaussianity.

So far, EoF cannot be exactly computed in the coupled harmonic oscillator system except in the ground state because of the non-Gaussian nature of exciting 
states\footnote{The R\'enyi-$\alpha$ entropies of few non-Gaussian states have been derived in \cite{Ilki-18}.}. 
As EoF and 
uncertainty exhibit similar behavior, as shown in  Eq. (\ref{two-vacuum}), the uncertainty may be used as a measure of entanglement after appropriate rescaling if EoF cannot be computed exactly. 
Therefore, in order to understand the entanglement more profoundly in the continuous variable system,
it is important to examine the uncertainty of the arbitrary excited states in the coupled 
harmonic oscillator system. 

In there any other similarity between EoF and uncertainty? EoF is believed to have the additivity property\cite{open05}, even though the property has still not been solved 
completely. For mixed states, EoF is generally defined by a convex-roof
method\cite{benn96,uhlmann99-1} as follows:
\begin{equation}
\label{roof}
{\cal E}_F (\rho) = \min \sum_i p_i {\cal E}_F (\rho_i),
\end{equation}
where the minimum is taken over all possible ensembles of pure states with $\sum_i p_i = 1$. Let $\rho^{(i)} \hspace{.2cm} (i=1,2)$ be two bipartite density
matrices, and $\rho = \rho^{(1)} \otimes \rho^{(2)}$. If we regard $\rho$ as a bipartite state, where $\rho^{(1)}$ and $\rho^{(2)}$ belong to each party, 
Eq. (\ref{roof}) guarantees ${\cal E}_F (\rho) \leq {\cal E}_F (\rho^{(1)}) + {\cal E}_F (\rho^{(2)})$. The additivity conjecture of EoF is that the 
equality always holds; this has been demonstrated through various examples  in \cite{additivity}. 
In this paper, we show that uncertainty in the coupled harmonic oscillator system also has a particular additive property, which we call the sum rule. 
We present this sum rule in the coupled harmonic oscillator system when the parameters are arbitrarily time-dependent. 

The paper is organized as follows. In Sec. II we derive $(\Delta x)^2$ and $(\Delta p)^2$ of arbitrary excited states by making use of explicit Wigner distribution  in the 
single harmonic oscillator system when the frequency is arbitrary time-dependent. It is shown that the time-dependence of frequency as well as energy level increase 
the uncertainty $\Delta x \Delta p$. In Sec. III we examine the uncertainties in the two-coupled harmonic oscillator system when the parameters are arbitrarily time-dependent. 
In this section we derive the uncertainties $(\Delta x)^2$ and $(\Delta p)^2$ of the first or second oscillator when two oscillators are at the arbitrary excited states. It is 
shown that $(\Delta x)^2$ and $(\Delta p)^2$ are just the arithmetic average of uncertainties of two single oscillators. We call this additive property as ''sum rule of quantum uncertainties''.
As a by-product, the purity function of the reduced state
is explicitly computed in this section by making use of the reduced Wigner distribution function. In Sec. IV we examine the uncertainties in the $N$-coupled harmonic oscillator system when 
 the parameters are arbitrarily time-dependent. It is shown that the arithmetic average property of uncertainties arising at $N=2$ is not generally maintained when $N \geq 3$. However, this additive
property is recovered when $(N-1)$ quantum numbers are equal. In Sec. V conclusion and further discussion are briefly given.

\section{Uncertainty for arbitrary excited state of single harmonic oscillator with arbitrary time-dependent frequency}

We start with a simple single harmonic oscillator Hamiltonian with arbitrary time-dependent frequency: $H_1 = \frac{p^2}{2} + \frac{1}{2} \omega^2 (t) x^2$. 
This simple model is important for studying  the squeezed states, which appear in various branches of physics, such as quantum 
optics\cite{walls-83,loudon-87,wu-87,schnabel-17} and cosmology\cite{grishchuk-90,grishchuk-93,einhorn-03,kiefer-07}.
The time-dependent  Schr\"{o}dinger equation (TDSE) of this system was examined in detail in \cite{lewis68,rhode89,lohe09,pires-19}. The linearly independent solutions $\psi_n (x, t) \hspace{.1cm} (n=0, 1, \cdots)$ are expressed 
in the following form\cite{lewis68,lohe09}:
\begin{equation}
\label{TDSE-1}
\psi_n (x, t) = e^{-i E_n \tau(t)} \frac{1}{\sqrt{2^n n!}} \left(\frac{\omega'}{\pi} \right)^{1/4}
 H_n (\sqrt{\omega'} x) e^{-\frac{v}{2} x^2}
\end{equation}
where $\omega' = \frac{\omega(0)}{b^2}$ and 
\begin{equation}
\label{TDSE-2}
v = \omega' - i \frac{\dot{b}}{b} 
\hspace{1.0cm}  E_n = \left( n + \frac{1}{2} \right) \omega(0)     \hspace{1.0cm}  \tau (t) = \int_0^t \frac{d s}{b^2 (s)}.
\end{equation}
In Eq. (\ref{TDSE-1}) $H_n (z)$ is the $n^{th}$-order Hermite polynomial and $b(t)$ satisfies the nonlinear Ermakov equation,
\begin{equation}
\label{ermakov-1}
\ddot{b} + \omega^2 (t) b = \frac{\omega^2 (0)}{b^3}
\end{equation}
with $b(0) = 1$ and $\dot{b} (0) = 0$. As shown in Eq. (\ref{TDSE-1}), $b(t)$ plays the role of scaling the frequency. Solutions of the Ermakov equation were discussed in \cite{lohe09,pinney50,gritsev-10,campo-16}. If $\omega(t)$ is time independent, $b(t)$ is simply one. 
If $\omega (t)$ is instantly changed as follows:
\begin{eqnarray}
\label{instant-1}
\omega (t) = \left\{                \begin{array}{cc}
                                               \omega_i  & \hspace{1.0cm}  t = 0   \\
                                               \omega_f  & \hspace{1.0cm}  t > 0,
                                               \end{array}            \right.
\end{eqnarray}
then $b(t)$ becomes
\begin{equation}
\label{scale-1}
b(t) = \sqrt{ \frac{\omega_f^2 - \omega_i^2}{2 \omega_f^2} \cos (2 \omega_f t) +  \frac{\omega_f^2 + \omega_i^2}{2 \omega_f^2}}.
\end{equation}
Of course, for a more general case of $\omega (t)$, the nonlinear Ermakov equation should be solved numerically or approximately. 

The $d$-dimensional  Wigner distribution function\cite{feyman72,noz91} is defined in terms of the phase space variables in the following form:
\begin{equation}
\label{dwigner}
W ({\bm x}, {\bm p}: t) = \frac{1}{\pi^d} \int d {\bm z} e^{-2 i {\bm p} \cdot {\bm z}} \Psi^* ({\bm x} + {\bm z}: t) 
 \Psi ({\bm x} - {\bm z}: t) 
\end{equation}
where ${\bm x} = (x_1, x_2, \cdots, x_d)$, ${\bm p} = (p_1, p_2, \cdots, p_d)$, and $ \Psi ({\bm r}: t)$ is a wave function of a given system.   
The Wigner distribution function is used to compute the expectation 
values. For example, the expectation value of $f(x_1, p_1)$ can be computed by 
\begin{equation}
\label{expectation}
\langle f(x_1, p_1) \rangle = \int d {\bm x} d {\bm p} f(x_1, p_1) W ({\bm x}, {\bm p}: t).
\end{equation}
Moreover, the Wigner distribution function has information on the substate of density matrix $\rho ({\bm x}, {\bm x'}: t) = \Psi ({\bm x}: t) \Psi^* ({\bm x'}: t)$. If $\rho_A (x_1, x_1': t) = \mbox{Tr}_{2, 3, \cdots, d} \rho ({\bm x}, {\bm x'}: t)$, the purity function of $\rho_A$ can be computed as 
\begin{equation}
\label{purity}
P_A (t) \equiv \mbox{Tr} \rho_A^2 = 2 \pi \int dx_1 dp_1 W^2 (x_1, p_1: t),
\end{equation}
where $W(x_1, p_1: t) = \int dx_2 \cdots dx_d dp_2 \cdots dp_d W ({\bm x}, {\bm p}: t)$.

To explicitly compute the Wigner distribution function of $H_1$, we set  $d=1$ and $\Psi = \psi_n (x, t)$ of Eq. (\ref{TDSE-1}) in Eq. (\ref{dwigner}). 
The integral in Eq. (\ref{dwigner}) can be computed by using\cite{integral}
\begin{eqnarray}
\label{main}
&&\int_{-\infty}^{\infty} dx e^{-p x^2 + 2 q x} H_m (a x + b) H_n (c x + d)                     \\     \nonumber
&&= \sqrt{\frac{\pi}{p}} e^{\frac{q^2}{p}} \sum_{k=0}^{\min (m,n)} \left(  \begin{array}{c} m \\  k  \end{array} \right)
\left(  \begin{array}{c} n \\  k  \end{array} \right) k! \left( 1 - \frac{a^2}{p} \right)^{\frac{m-k}{2}} \left( 1 - \frac{c^2}{p} \right)^{\frac{n-k}{2}}
\left( \frac{2 a c}{p} \right)^k                                                                                   \\    \nonumber         
&&  \hspace{3.0cm} \times H_{m-k} \left(\frac{b + \frac{a q }{p}}{\sqrt{1 - \frac{a^2}{p}}} \right)  H_{n-k} \left(\frac{d + \frac{c q }{p}}{\sqrt{1 - \frac{c^2}{p}}} \right).
\end{eqnarray}
Then, the Wigner distribution function for $H_1$ can be written as follows:
\begin{eqnarray}
\label{wigner-H1}
&&W_n (x, p: t) = \frac{1}{\pi} \exp \left[ - \omega' x^2 - \frac{1}{\omega'} \left( p + \frac{\dot{b}}{b} x \right)^2 \right]     \\    \nonumber
&&\hspace{3.0cm} \times\sum_{k=0}^n \left(  \begin{array}{c} n \\  k  \end{array} \right) (-1)^k \frac{2^{n-k}}{(n-k)!} 
\left[ \omega' x^2 + \frac{1}{\omega'} \left( p + \frac{\dot{b}}{b} x \right)^2 \right]^{n - k}                     \\   \nonumber
&& = \frac{1}{n! \pi} \exp \left[ - \omega' x^2 - \frac{1}{\omega'} \left( p + \frac{\dot{b}}{b} x \right)^2 \right]
U \left(-n, 1, 2 \left[ \omega' x^2 + \frac{1}{\omega'} \left( p + \frac{\dot{b}}{b} x \right)^2 \right] \right),
\end{eqnarray}
where $U(a, b, z)$ is a confluent hypergeometric function. 
It is straightforward to show that $\int dx dp W_n (x, p: t) = 2 \pi \int dx dp W_n^2 (x, p: t) = 1$, which guarantees $\psi_n (x, t)$ is the normalized pure state. 
By using the Wigner distribution function, it is straightforward to show that for non-negative integer $m$,
$\langle x^{2 m+1} \rangle = \langle p^{2m + 1} \rangle = 0$ and 
\begin{eqnarray}
\label{single-expect}
&&\langle x^{2 m} \rangle = \frac{2^n (m + n)!}{m! n! \sqrt{\pi} \omega'^m} \Gamma \left( \frac{2 m + 1}{2} \right)   {_2F_1} \left( -n, -n: -n - m: 1 / 2 \right)                                                                        \\     \nonumber
&&\langle p^{2 m} \rangle = \frac{2^n (m + n)!}{m! n! \sqrt{\pi}} \Gamma \left( \frac{2 m + 1}{2} \right) 
\left[ \omega' + \frac{1}{\omega'} \left(  \frac{\dot{b}}{b} \right)^2 \right]^m   {_2F_1} \left( -n, -n: -n - m: 1 / 2 \right),
\end{eqnarray}
where $\Gamma (z)$ and ${_2F_1} (a, b: c: z)$ are gamma and hypergeometric functions. Thus, the uncertainties for $x$ and $p$ are 
\begin{equation}
\label{uncertainty-1}
\left(\Delta x\right)^2 = \frac{n + \frac{1}{2}}{\omega'}   \hspace{1.0cm} 
\left(\Delta p\right)^2 = \left( n + \frac{1}{2} \right) \left[\omega' + \frac{1}{\omega'} \left( \frac{\dot{b}}{b} \right)^2 \right],
\end{equation}
which yield an uncertainty relation
\begin{equation}
\label{uncertainty-2}
\left( \Delta x \Delta p \right)^2 =  \left( n + \frac{1}{2} \right)^2 \left[1 + \frac{1}{\omega'^2} \left( \frac{\dot{b}}{b} \right)^2 \right].
\end{equation}
Thus, the time-dependence of $\omega$ as well as energy level $n$ increase the uncertainty $\Delta x \Delta p$. 

\section{Uncertainty for arbitrary excited state of two-coupled harmonic oscillator system with arbitrary time-dependent parameters}

Now, let us consider the Hamiltonian (\ref{hamil-2}) again when $k_0$ and $J$ are arbitrarily time dependent. It is not difficult to show that the 
Hamiltonian is diagonalized by introducing normal coordinates  $y_1 =  (x_1 + x_2) / \sqrt{2}$ and $y_2 = (-x_1 + x_2) / \sqrt{2}$, and their conjugate momenta $\pi_1$
and $\pi_2$ with normal mode frequencies $\omega_1 = \sqrt{k_0}$ and $\omega_2 = \sqrt{k_0 + 2 J}$, respectively. If two oscillators are in the
$n^{th}$ and $m^{th}$ states, we will show in the following that the uncertainties for $x_j$ and $p_j$  ($j = 1, 2$) are just the arithmetic mean of two single oscillators; 
that is, 
\begin{eqnarray}
\label{two-coupled}
&&\left(\Delta x_1\right)^2 = \left(\Delta x_2\right)^2 = \frac{1}{2} \left[ \frac{2 n + 1}{2 \omega_1'} + \frac{2 m + 1}{2 \omega_2'} \right]    
                                                                                                                                                                           \\   \nonumber
&&\left(\Delta p_1\right)^2 = \left(\Delta p_2\right)^2 = \frac{1}{2} \left[ \frac{2 n + 1}{2} \left\{ \omega_1' + \frac{1}{\omega_1'} \left( \frac{\dot{b}_1}{b_1} \right)^2 \right\} + 
\frac{2 m + 1}{2} \left\{ \omega_2' + \frac{1}{\omega_2'} \left( \frac{\dot{b}_2}{b_2} \right)^2 \right\}  \right],
\end{eqnarray}
where $\omega_j' = \omega_j (0) / b_j^2$ ($j=1,2$), and $b_j$ satisfy their own nonlinear Ermakov equations,  $\ddot{b}_j + \omega_j^2 (t) b_j = \frac{\omega_j^2 (0)}{b_j^2}$  with $\dot{b}_j (0) = 0$ and $b_j (0) = 1$. We will call this arithmetic 
average additivity as ``sum rule of quantum uncertainty''.

To prove  Eq. (\ref{two-coupled}), we start with solutions of TDSE for $H_2$ in terms of $y_j$, which is 
\begin{eqnarray}
\label{TDSE-3}
&&\psi_{n,m} (x_1, x_2: t) = \frac{1}{\sqrt{2^{(n+m)} n! m!}} \left( \frac{\omega_1' \omega_2'}{\pi^2} \right)^{1/4} H_n (\sqrt{\omega_1'} y_1) 
 H_m (\sqrt{\omega_2'} y_2)                                                        \\    \nonumber
&& \hspace{3.0cm}  \times \exp \left[ -i (E_{n,1} \tau_1 + E_{m,2} \tau_2) - \frac{1}{2} \left(v_1 y_1^2 + v_2 y_2^2 \right)  \right],
\end{eqnarray}
where $E_{m,j} = \left( m + \frac{1}{2} \right) \omega_j (0)$, $\tau_j = \int_0^t \frac{ds}{b_j^2(s)}$, and 
$v_j = \omega_j' - i \frac{\dot{b}_j}{b_j}$. Now, let us compute the Wigner distribution functions of the $H_2$ system by setting
$\Psi ({\bm x}: t) = \psi_{n,m} (x_1, x_2: t)$ in Eq. (\ref{dwigner}).  If we change  Eq. (\ref{TDSE-3}) into the original phase space variables 
$x_j$ and $p_j$, and insert them into Eq. (\ref{dwigner}), the computation of the Wigner distribution function is highly complicated. However,  this difficulty can be avoided. 
Since $y_j$ are orthogonal normal modes, they preserve the inner product and $2$-dimensional volume elements. Thus, the Wigner distribution function for 
$H_2$ is simply reduced to 
\begin{equation}
\label{wigner-H2}
W_{n,m} (x_1, x_2: p_1, p_2: t) = W_n (y_1, \pi_1: t) \bigg|_{\omega' \rightarrow \omega_1', b \rightarrow b_1}
\times  W_m (y_2, \pi_2: t) \bigg|_{ \omega' \rightarrow  \omega_2' , b \rightarrow b_2  },
\end{equation}
where $W_n$ is a Wigner distribution function of a single harmonic oscillator given in Eq. (\ref{wigner-H1}). 

At this stage we want to digress little bit. Sometimes, we need to derive the lower-dimensional reduced Wigner distribution function to explore the properties of 
the reduced quantum state. Although we can compute the $2$-dimensional Wigner distribution function quickly by using the normal mode,
the derivation of the reduced $1$-dimensional Wigner distribution function is very complicated problem. For example, let us consider 
$W_{n,m} (x_1, p_1: t) \equiv \int dx_2 dp_2 W_{n,m} (x_1, x_2: p_1, p_2: t)$; here, the difficulty arises because $dx_2 dp_2$ is not invariant measure 
in the normal modes. Thus, we should compute the reduced Wigner distribution function by using the original coordinates and conjugate momenta. 
After long and tedious calculation, it is possible to show that
\begin{eqnarray}
\label{reduced-wigner-1}
&&W_{n,m} (x_1, p_1: t) = \frac{\sqrt{4 \omega_1' \omega_2'}}{\pi} \sum_{k=0}^n \sum_{\ell = 0}^m  
\left(  \begin{array}{c} n \\  k  \end{array} \right)    \left(  \begin{array}{c} m \\  \ell  \end{array} \right)
\frac{(-1)^{k + \ell}}{(n - k)! (m - \ell)!} 2^{(n + m) - (k + \ell)}                                              \\          \nonumber
&&  \times  \left( - \frac{\partial}{\partial \mu_1} \right)^{n-k}  \left( - \frac{\partial}{\partial \mu_2} \right)^{m-\ell} 
\frac{1}{\sqrt{\Omega (\mu_1, \mu_2: t)}} \exp \left[ - 2 \frac{\Theta (x_1, p_1: \mu_1, \mu_2: t)}{\Omega (\mu_1, \mu_2: t)} \right]
\Bigg|_{\mu_1 = \mu_2 = 1},
\end{eqnarray}
where 
\begin{eqnarray}
\label{reduced-wigner-2}
&&\Omega (\mu_1, \mu_2: t) = \omega_1'  \omega_2' (\mu_1^2 + \mu_2^2) + \left[ \omega_1'^2 + \omega_2'^2 + 
\left(\frac{\dot{b}_1}{b_1} - \frac{\dot{b}_2}{b_2} \right)^2 \right] \mu_1 \mu_2                                          \\    \nonumber
&&\Theta (x_1, p_1: \mu_1, \mu_2: t) = \omega_1' \left[ \omega_2'^2 x_1^2 + \left( p_1 +  \frac{\dot{b}_2}{b_2} x_1 \right)^2 \right] \mu_1^2 
\mu_2 + \omega_2' \left[ \omega_1'^2 x_1^2 + \left( p_1 +  \frac{\dot{b}_1}{b_1} x_1 \right)^2 \right] \mu_1 \mu_2^2.
\end{eqnarray}
Thus, the reduced Wigner distribution function for $n = m = 0$ is easily computed by 
\begin{equation}
\label{reduced-wigner-3}
W_{0, 0} (x_1, p_1: t) = \frac{1}{\pi} \sqrt{\frac{4 \omega_1' \omega_2'}{\Omega(1, 1: t)}} e^{-2 \Theta (x_1, p_1: 1, 1,: t) / \Omega(1, 1: t)}.
\end{equation}
The purity function of the $A$-oscillator is defined as $P_{n, m}^A (t) = \mbox{tr} \rho_{n, m}^2 (x_1, x_1': t)$, where $\rho_{n, m} (x_1, x_1': t)$ is an effective state of the $A$-oscillator
derived by taking a partial trace to $\rho_{n,m} (x_1, x_2: x_1', x_2': t) = \psi_{n, m} (x_1, x_2: t) \psi_{n,m}^* (x_1', x_2' :t)$ over $B$-oscillator. Then, 
$P_{0,0}^A (t)$ can be computed from $W_{0,0} (x_1, p_1: t)$ as follows: 
\begin{equation}
\label{purity-1}
P_{0, 0}^A (t) = 2 \pi \int dx_1 dp_1 W_{0, 0}^2(x_1, p_1: t) = 2 \sqrt{z}
\end{equation}
where $z = \omega_1' \omega_2' / \Omega(1, 1: t)$. From Eq. (\ref{reduced-wigner-1}) one can show directly 
$\int dx_1 dp_1 W_{n,m}(x_1, p_1: t) = 1$ by making use of simple binomial formula. Furthermore, it is possible to show that
\begin{eqnarray}
\label{purity-mn}
&& 2 \pi \int dx_1 dp_1 W_{m,n}^2 (x_1, p_1: t)                                                                
= 4 \sqrt{\omega_1' \omega_2'} \sum_{k,k' = 0}^n \sum_{\ell,\ell' = 0}^m                                             
\left(  \begin{array}{c} n \\  k  \end{array} \right)   \left(  \begin{array}{c} n \\  k'  \end{array} \right)              \\    \nonumber
&&\times \left(  \begin{array}{c} m \\  \ell  \end{array} \right)  \left(  \begin{array}{c} m \\  \ell'  \end{array} \right)
 \frac{(-1)^{k+k'+\ell+\ell'}}{(n-k)! (n-k')! (m-\ell)! (m- \ell')!}                                     
 2^{2(n+m) - (k+k'+\ell+\ell')}                                                                                                                                     \\   \nonumber
&&  \left( - \frac{\partial}{\partial \mu_1} \right)^{n-k}
  \left( - \frac{\partial}{\partial \nu_1} \right)^{n-k'}  \left( - \frac{\partial}{\partial \mu_2} \right)^{m-\ell}
   \left( - \frac{\partial}{\partial \nu_2} \right)^{m-\ell'} \frac{1}{\sqrt{\Gamma (\mu_1, \mu_2: \nu_1, \nu_2)}}
   \Bigg|_{\mu_1 = \mu_2 = \nu_1 = \nu_2 = 1}
\end{eqnarray}
where
\begin{eqnarray}
\label{Big-Gamma}
&&\Gamma(\mu_1, \mu_2: \nu_1, \nu_2 ) = \omega_1' \omega_2' \left[ \mu_1^2 \nu_1^2 (\mu_2 + \nu_2)^2 + \mu_2^2 \nu_2^2 (\mu_1 + \nu_1)^2 \right]                                                                         \\     \nonumber
&&\hspace{2.0cm} + \mu_1 \mu_2 \nu_1 \nu_2 (\mu_1 + \nu_1) (\mu_2 + \nu_2) \left[ \omega_1'^2 + \omega_2'^2 + \left( \frac{\dot{b}_1}{b_1} - \frac{\dot{b}_2}{b_2} \right)^2   \right].
\end{eqnarray}
If we define the ratios
\begin{equation}
\label{ratio}
\gamma_n = \frac{P_{n,0}^A (t)}{P_{0, 0}^A (t)}      \hspace{1.0cm}
\delta_n = \frac{P_{n,n}^A (t)}{P_{0, 0}^A (t)} ,
\end{equation}
they are summarized at Table I.  We expect that $\gamma_n$ and $\delta_n$ decrease with increasing $n$ because more excited states seem to be 
more mixed. 

\begin{center}
\begin{tabular}{c|c|c} \hline \hline
$n$ & $\gamma_n$     & $\delta_n$                             \\  \hline \hline
$1$ &  $\frac{1}{4} (3 - 4 z)$    &   $\frac{1}{16} (9 - 40 z + 144 z^2)$               \\    
$2$ & $\frac{1}{64} (41 - 104 z + 144 z^2)$      &    $\frac{1}{4096} (1681 - 19344 z + 256608 z^2 - 1440000 z^3 + 2822400 z^4)$              \\  
$3 $  &  $\frac{1}{256} (147 - 540 z + 1488 z^2 - 1600z^3)$    &   too long              \\     \hline
\end{tabular}

\vspace{0.3cm}
Table I: The ratios $\gamma_n$ and $\delta_n$ for $n=1, 2, 3$
\end{center}
\vspace{0.5cm}

\begin{figure}[ht!]
\begin{center}
\includegraphics[height=5.0cm]{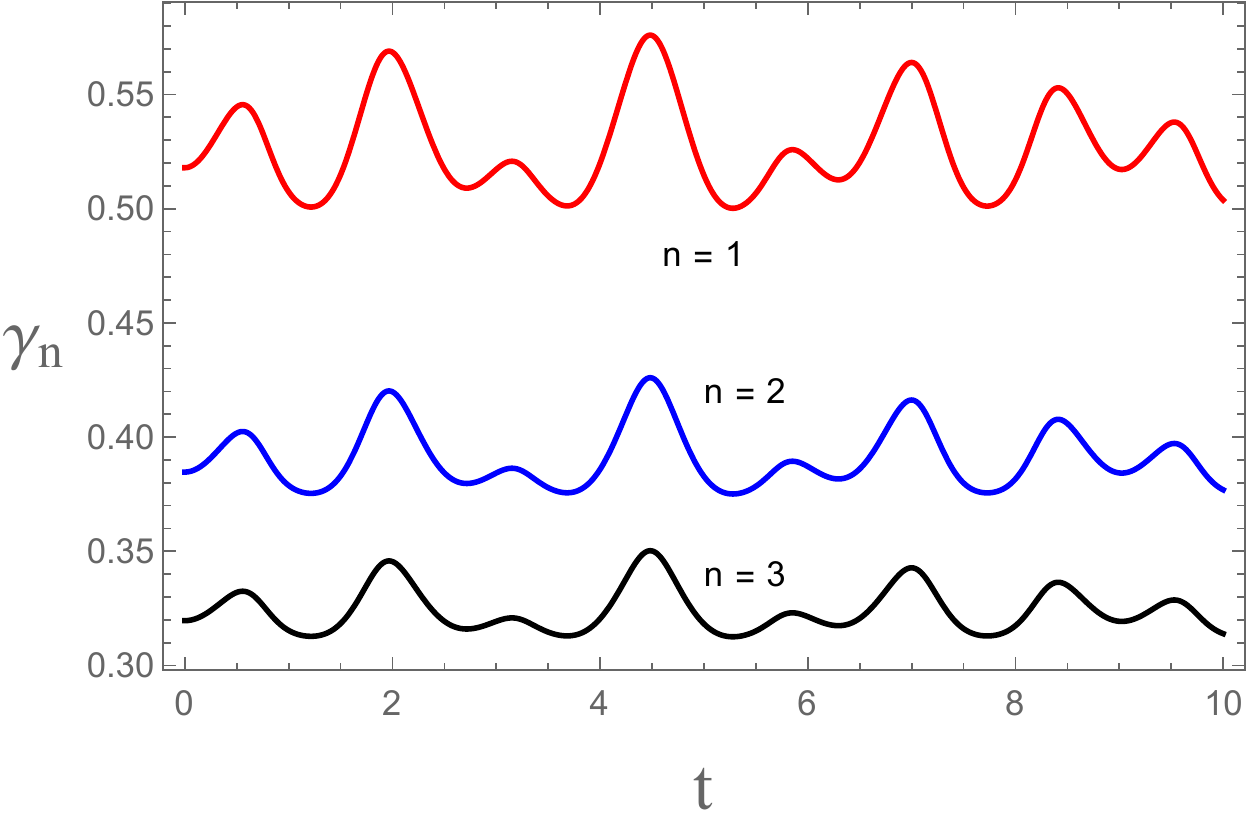} \hspace{0.5cm}
\includegraphics[height=5.0cm]{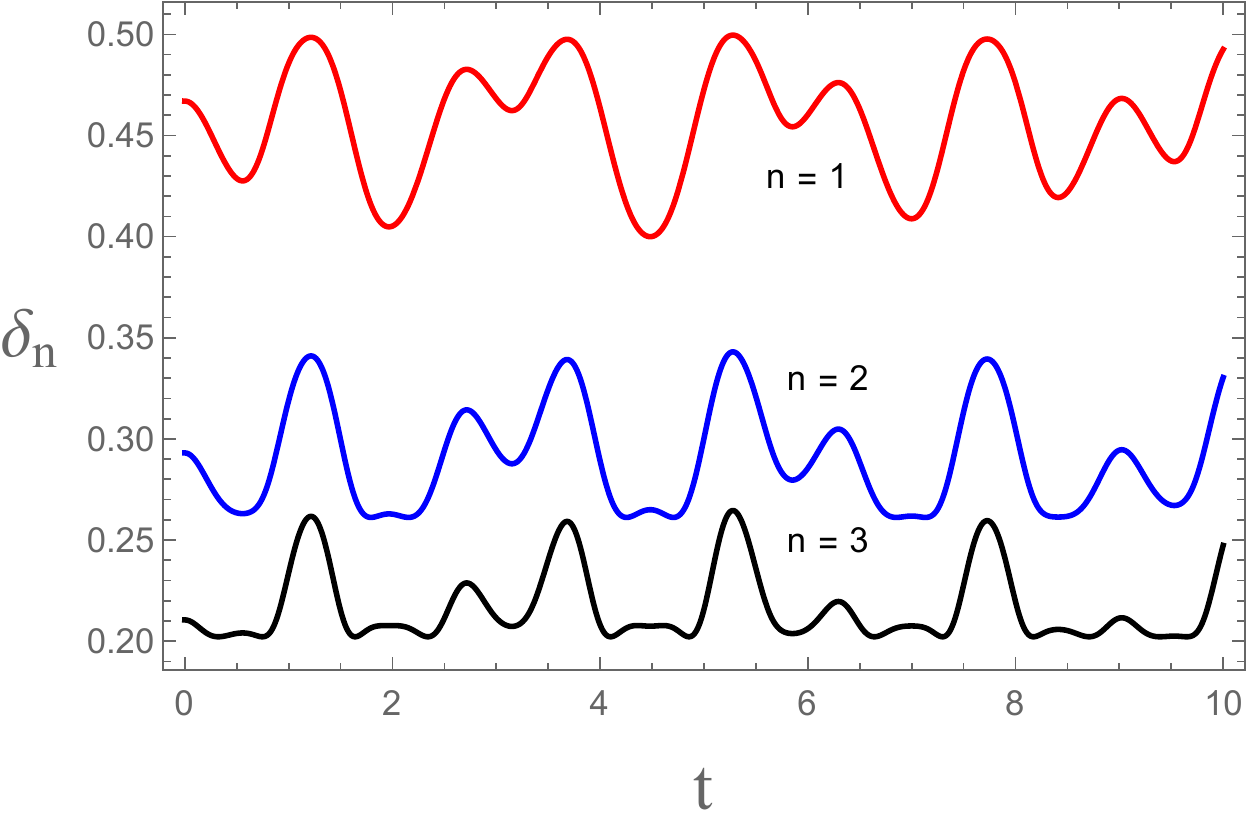}

\caption[fig1]{(Color online) The time dependence of the ratios (a) $\gamma_n$ and (b) $\delta_n$
when $k_0 (0) = J (0) = 1$ and $k_0 (t) = J(t) = 2 \hspace{.2cm} (t>0)$.
As expected, the figures exhibit that  the effective states for $A$-oscillator are more and more mixed with increasing $n$. }
\end{center}
\end{figure}
The time dependence of $\gamma_n$ and $\delta_n$ is plotted in Fig. 1(a) and Fig. 1(b) when $k_0 (0) = J (0) = 1$ and $k_0 (t) = J(t) = 2 \hspace{.2cm} (t>0)$.
As expected, the figures exhibit that  the effective states for the $A$-oscillator is more and more mixed with increasing $n$. 
Remarkably, this figure shows that the reduced state of $\rho_{n,n}$ is more mixed than that of $\rho_{n,0}$.


Now, let us return to discuss about the uncertainties. From Eq. (\ref{wigner-H2}), it is easy to show that 
$\langle y_j^{2 m + 1} \rangle = \langle \pi_j^{2 m + 1} \rangle = 0$, and  $\langle y_j^{2 m}\rangle$ and  $\langle \pi_j^{2 m} \rangle$ are equal to 
$\langle x^{2 m} \rangle$ and $\langle p^{2 m} \rangle$, respectively, in Eq. (\ref{single-expect}) with changing $\omega' \rightarrow \omega_j'$ and 
$b \rightarrow b_j$. Accordingly, by using this fact and the normal modes, it is easy to prove Eq. (\ref{two-coupled}). 

\section{Uncertainty for arbitrary excited state of $N$-coupled harmonic oscillator system with arbitrary time-dependent parameters}

To check whether the property of arithmetic average for uncertainties is maintained in a multi-coupled harmonic oscillator system or not, we first consider a three-coupled 
harmonic oscillator system with the following  Hamiltonian: 
\begin{equation}
\label{hamil-3}
H_3 = \frac{1}{2} (p_1^2 + p_2^2 + p_3^2) + \frac{1}{2}\left[ k_0 (t) (x_1^2 + x_2^2 + x_3^2) + J (t) \left\{ (x_1 - x_2)^2 + (x_1 - x_3)^2 + (x_2 - x_3)^2 \right\} \right].
\end{equation}
The normal mode coordinates of $H_3$ is $y_1 = (x_1 + x_2 + x_3) / \sqrt{3}$, $y_2 = (x_1 - x_2) / \sqrt{2}$, and $y_3 = (x_1 + x_2 - 2 x_3) / \sqrt{6}$
with normal mode frequencies $\omega_1 = \sqrt{k_0}$ and $\omega_2 = \omega_3 = \sqrt{k_0 + 3 J} \equiv \omega$. If three oscillators 
are in the $n^{th}$, $m^{th}$, and $\ell^{th}$ states,  the $3$-dimensional Wigner distribution function can be computed as follows:
\begin{equation}
\label{wigner-H3}
W_{n,m} (x_1, x_2, x_3: p_1, p_2, p_3: t) = W_n (y_1, \pi_1: t) \bigg|_{\omega' \rightarrow \omega_1', b \rightarrow b_1}
\times  W_m (y_2, \pi_2: t)  \times W_{\ell} (y_3, \pi_3: t)
\end{equation}
where $\pi_j$ represent the conjugate momenta of $y_j$ and $W_n$ is the Wigner distribution function of the single harmonic oscillator given in Eq. (\ref{wigner-H1}). 
Of course, $b_1(t)$ and $b(t)$ are solutions for Ermakov equations for $\omega_1$ and $\omega$, and $\omega_1' = \omega_1 (0) / b_1^2 (t)$ and 
$\omega' = \omega (0) / b^2 (t)$. Thus, Eq. (\ref{uncertainty-1}) and Wigner distribution function (\ref{wigner-H3}) imply
\begin{eqnarray}
\label{revise-1}
&&\left(\Delta y_1\right)^2 = \frac{n + \frac{1}{2}}{\omega_1'}   \hspace{1.0cm} 
\left(\Delta \pi_1 \right)^2 = \left( n + \frac{1}{2} \right) \left[\omega_1' + \frac{1}{\omega_1'} \left( \frac{\dot{b_1}}{b_1} \right)^2 \right]                   \\    \nonumber
&&\left(\Delta y_2\right)^2 = \frac{m + \frac{1}{2}}{\omega'}   \hspace{1.0cm} 
\left(\Delta \pi_2 \right)^2 = \left( m + \frac{1}{2} \right) \left[\omega' + \frac{1}{\omega'} \left( \frac{\dot{b}}{b} \right)^2 \right]                           \\    \nonumber
&&\left(\Delta y_3\right)^2 = \frac{\ell + \frac{1}{2}}{\omega'}   \hspace{1.0cm} 
\left(\Delta \pi_3 \right)^2 = \left( \ell + \frac{1}{2} \right) \left[\omega' + \frac{1}{\omega'} \left( \frac{\dot{b}}{b} \right)^2 \right].
\end{eqnarray}
Then, it is straightforward to show 
\begin{eqnarray}
\label{three-coupled}
&& \left(\Delta x_1 \right)^2 = \left( \Delta x_2 \right)^2 = \frac{1}{3} \left[ \frac{2 n + 1}{2 \omega_1'} + \frac{3 (2 m + 1) + (2 \ell + 1)}{4 \omega'} \right]                                                                                                    \\    \nonumber
&& \left( \Delta x_3 \right)^2 = \frac{1}{3} \left[ \frac{2 n + 1}{2 \omega_1'} + 2 \frac{2 \ell + 1}{2 \omega'} \right]           \\    \nonumber
&& \left( \Delta p_1 \right)^2 = \left( \Delta p_2 \right)^2 = \frac{1}{3} \left[ \frac{2 n + 1}{2} \left\{ \omega_1' + \frac{1}{\omega_1'} \left( \frac{\dot{b}_1}{b_1} \right)^2  \right\} + \frac{3 (2 m + 1) + (2 \ell + 1)}{4} \left\{ \omega' + \frac{1}{\omega'} \left( \frac{\dot{b}}{b} \right)^2  \right\}   \right]                                  \\      \nonumber
&& \left( \Delta p_3 \right)^2 = \frac{1}{3} \left[ \frac{2 n + 1}{2} \left\{ \omega_1' + \frac{1}{\omega_1'} \left( \frac{\dot{b}_1}{b_1} \right)^2  \right\} + 2 \frac{2 \ell + 1}{2} \left\{ \omega' + \frac{1}{\omega'} \left( \frac{\dot{b}}{b} \right)^2  \right\}   \right].
\end{eqnarray}
Thus, the property of the arithmetic average in uncertainties is not maintained when $N=3$. However, this property is recovered when $m = \ell$. 

Finally, let us consider the $N$-coupled harmonic oscillator system with the following Hamiltonian:
\begin{equation}
\label{hamil-N}
H_N = \frac{1}{2} \sum_{i=1}^N p_i^2 + \frac{1}{2} \left[ k_0 (t) \sum_{i=1}^N x_i^2 + J(t) \sum_{i<j}^N (x_i - x_j)^2 \right].
\end{equation}
This system is diagonalized by introducing the normal mode coordinates $y_1 = (x_1 + x_2 + \cdots + x_N) / \sqrt{N}$ and $y_j = (x_1 + x_2 + \cdots + x_{j-1} - (j-1) x_j) / \sqrt{j (j-1)} \hspace{.2cm} (j = 2, 3, \cdots, N)$ with normal mode frequencies $\omega_1 = \sqrt{k_0}$ and $\omega_2 = \omega_3 = \cdots = \omega_N = \sqrt{k_0 + N J} \equiv \omega$. If $N$ oscillators are in the $n_1^{th}, n_2^{th}, \cdots, n_N^{th}$ states, the $N$-dimensional 
Wigner distribution function can be written as follows:
\begin{equation}
\label{wigner-H-N}
W_{n_1, n_2, \cdots, n_N} ({\bm x}, {\bm p}: t) = W_{n_1} (y_1, \pi_1: t) \bigg|_{\omega' \rightarrow \omega_1', b \rightarrow b_1}
\times  \prod_{j=2}^N W_{n_j} (y_j, \pi_j: t),
\end{equation} 
where $\pi_j$ represent the conjugate momenta of $y_j$ and $W_n$ is the Wigner distribution function of the single harmonic oscillator given in Eq. (\ref{wigner-H1}). 
Then, it is straightforward to show that
\begin{eqnarray}
\label{N-coupled}
&&\left( \Delta x_j \right)^2 = \frac{1}{N} \left[ \frac{2 n_1+1}{2 \omega_1'} + \frac{1}{2 \omega'} \left\{ \frac{2 N (j - 1)}{j} n_j + 2 N \sum_{k = j+1}^N \frac{n_k}{k (k - 1)} + (N-1) \right\} \right]                                       \\    \nonumber
&&\left( \Delta p_j \right)^2 = \frac{1}{N}  \Bigg[ \frac{2 n_1+1}{2} \left\{ \omega_1' + \frac{1}{\omega_1'} \left( \frac{\dot{b}_1}{b_1} \right)^2  \right\}                                                                                                                                      \\    \nonumber
 &&\hspace{2.0cm}+ \frac{1}{2} \left\{ \frac{2 N (j - 1)}{j} n_j + 2 N \sum_{k = j+1}^N \frac{n_k}{k (k - 1)} + (N-1) \right\} 
 \left\{ \omega' + \frac{1}{\omega'} \left( \frac{\dot{b}}{b} \right)^2  \right\} \Bigg].
\end{eqnarray}
Eq. (\ref{N-coupled}) can be shown to reproduces Eq. (\ref{two-coupled}) and Eq. (\ref{three-coupled}) when $N=2$ and $N=3$ if the quantum numbers 
$n_1$, $n_2$, and $n_3$ are replaced by $n$, $m$, and $\ell$, respectively. If $n_2 = n_3 = \cdots = n_N$, one can show that $\left( \Delta x_j \right)^2$ and 
$\left( \Delta p_j \right)^2$ are independent of $j$ and they are just arithmetic average of uncertainties for each oscillator. 

\section{Conclusions}

In this paper we computed the uncertainties of $(\Delta x)^2$ and $(\Delta p)^2$ analytically in an $N$-coupled harmonic oscillator system. When $N = 2$, 
these uncertainties are just the arithmetic average of uncertainties of two single harmonic oscillators. We call this 
property as ``sum rule of quantum uncertainty''. However, this additive property is not generally 
maintained when $N \geq 3$ but is recovered in an $N$-coupled oscillator system only when $(N-1)$ quantum numbers are equal. 

Our calculation can be generalized to a more general case. For example, let us consider the following Hamiltonian 
\begin{eqnarray}
\label{H3-tilde}
&&\tilde{H}_3 = \frac{1}{2} \left( p_1^2 + p_2^2 + p_3^2 \right) + \frac{1}{2} \bigg[ k_0 (t) \left( x_1^2 + x_2^2 + x_3^2 \right) + J_{12} (t) (x_1 - x_2)^2
                                                                                                                                                                             \\    \nonumber
&&\hspace{6.0cm} + J_{13} (t) (x_1 - x_3)^2 + J_{23} (t) (x_2 - x_3)^2 \bigg].
\end{eqnarray}
In this case, the normal mode coordinates become 
\begin{eqnarray}
\label{normal-H3-tilde}
&& y_1 = \frac{1}{\sqrt{3}} (x_1 + x_2 + x_3)                                                      \\     \nonumber
&& y_+ = A_+ (-J_{12} + J_{23} - \zeta) x_1 + A_+ (J_{12} - J_{13} + \zeta) x_2 + A_+ (J_{13} - J_{23}) x_3                       \\    \nonumber
&& y_- = A_- (-J_{12} + J_{23} + \zeta) x_1 + A_- (J_{12} - J_{13} - \zeta) x_2 + A_- (J_{13} - J_{23}) x_3
\end{eqnarray}
with $\zeta = \sqrt{J_{12}^2 + J_{13}^2 + J_{23}^2 - (J_{12} J_{13} + J_{12} J_{23} + J_{13} J_{23})}$ and 
\begin{equation}
\label{Apm}
A_{\pm} = \frac{1}{J_{13} - J_{23}}\sqrt{\frac{2 \zeta \pm (J_{13} + J_{23} - 2 J_{12})}{6 \zeta}}.
\end{equation}
Moreover, the normal mode frequencies are given by $\omega_1 = \sqrt{k_0}$ and $\omega_{\pm} = \sqrt{k_0 + J_{12} + J_{13} + J_{23} \pm \zeta}$. 
If the three oscillators are in the $n^{th}$, $m^{th}$, and $\ell^{th}$ exciting states, our procedure yields
\begin{eqnarray}
\label{uncertainty-H3-tilde}
&&\left( \Delta x_1 \right)^2                                                          
= \frac{1}{3} \frac{2 n + 1}{2 \omega_1'} + A_+^2 u_-^2 \frac{2 m + 1}{2 \omega_+'} + 
A_-^2 u_+^2 \frac{2 \ell + 1}{2 \omega_-'}                                     \\    \nonumber
&&\left( \Delta x_2 \right)^2                                                          
= \frac{1}{3} \frac{2 n + 1}{2 \omega_1'} + A_+^2 v_+^2 \frac{2 m + 1}{2 \omega_+'} + 
A_-^2 v_-^2 \frac{2 \ell + 1}{2 \omega_-'}                                     \\    \nonumber
&&\left( \Delta x_3 \right)^2  = \frac{1}{3} \frac{2 n + 1}{2 \omega_1'} 
+ (J_{13} - J_{23})^2 \left[ A_+^2 \frac{2 m + 1}{2 \omega_+'} + A_-^2 \frac{2 \ell + 1}{2 \omega_-'} \right],
\end{eqnarray}
where $u_{\pm} = -J_{12} + J_{23} \pm \zeta$, $v_{\pm} = J_{12} - J_{13} \pm \zeta$, and 
$\omega_{j}' = \omega_j / b_j^2 (t) \hspace{.2cm} (j = 1, \pm)$. Of course $b_j$ are the scaling factors of $\omega_j$. Similarly, the uncertainties
$\left( \Delta p_j \right)^2$ can be computed explicitly by following the same procedure. 

We do not know whether or not the sum rule of quantum uncertainty arising at $N = 2$ is realized in other continuous variable systems such as $1/x$-potential system.
Also, we do not clearly understand whether or not the sum rule of uncertainty may have some implication on the additivity of entanglement. We hope to explore 
these issues in the future. 

Quantum information processing with continuous variables has attracted considerable attention from both theoretical and experimental aspects\cite{braunstein-2005,adesso-2014}.
Quantum uncertainties are closely connected to the inseparability criterion of a continuous-variable quantum system\cite{criterion,simon-2000}.
Furthermore, the distillation protocols to a maximally entangled state have already been suggested 
in Duan et al.\cite{duan-99-p} and Giedke et al.\cite{giedke-2000}.
We hope that our results on the explicit expressions of uncertainties may give valuable insight into the problem of 
continuous-variable quantum information processing.


\end{document}